\documentclass{article}

\usepackage{arxiv}

\usepackage[utf8]{inputenc} % allow utf-8 input
\usepackage[T1]{fontenc}    % use 8-bit T1 fonts
\usepackage{hyperref}       % hyperlinks
\usepackage{url}            % simple URL typesetting
\usepackage{booktabs}       % professional-quality tables
\usepackage{amsfonts}       % blackboard math symbols
\usepackage{nicefrac}       % compact symbols for 1/2, etc.
\usepackage{microtype}      % microtypography
\usepackage{lipsum}		% Can be removed after putting your text content
\usepackage{graphicx}
\usepackage{doi}
\usepackage{indentfirst}
\usepackage{amsmath}

\title{Stock Index Prediction using Cointegration test and Quantile Loss
}

\author{{\hspace{1mm}Jaeyoung Cheong}\thanks{musicjae@yonsei.ac.kr} \\
	Department of Philosophy\\
	Yonsei University\\
	Seoul, Republic of Korea \\
	\texttt{musicjae@yonsei.ac.kr} \\
	\And
	{\hspace{1mm}Heejoon Lee}\thanks{peter@thecodit.com} \\
	CODIT Corp.\\
	Seoul, Republic of Korea \\
	\texttt{peter@thecodit.com} \\
	\And
	{\hspace{1mm}Minjung Kang} \\
	Department of International Business\\
	Chungbuk National University\\
	Daejeon, Republic of Korea \\
	\texttt{miinkang.dev@gmail.com} \\

}

\hypersetup{
pdftitle={A template for the arxiv style},
pdfsubject={q-bio.NC, q-bio.QM},
pdfauthor={David S.~Hippocampus, Elias D.~Striatum},
pdfkeywords={First keyword, Second keyword, More},
}

\begin{document}
\maketitle

\begin{abstract}
Recent researches on stock prediction using deep learning methods has been actively studied. This is the task to predict the movement of stock prices in the future based on historical trends. The approach to predicting the movement based solely on the pattern of the historical movement of it on charts, not on fundamental values, is called the Technical Analysis, which can be divided into univariate and multivariate methods in the regression task. According to the latter approach, it is important to select different factors well as inputs to enhance the performance of the model. Moreover, its performance can depend on which loss is used to train the model. However, most studies tend to focus on building the structures of models, not on how to select informative factors as inputs to train them. In this paper, we propose a method that can get better performance in terms of returns when selecting informative factors using the cointegration test and learning the model using quantile loss. We compare the two RNN variants with quantile loss with only five factors obtained through the cointegration test among the entire 15 stock index factors collected in the experiment. The Cumulative return and Sharpe ratio were used to evaluate the performance of trained models. Our experimental results show that our proposed method outperforms the other conventional approaches.
\end{abstract}

% keywords can be removed
\keywords{Stock Prediction \and Quantile loss \and Backtesting \and Cointegration test \and Deep Learing}

\section{Introduction}
Recently, it has been actively studied to predict stock index prices using deep learning \cite{ozbayoglu2020deep}, \cite{jiang2020applications}. This task is to predict stock index prices in the future based on historical stock index prices. Traditionally, the approach of stock index prediction is divided into Fundamental Analysis and Technical analysis. The former is the method to evaluate the value of a company by analyzing its financial statements, soundness, etc. On the other hand, the latter is the method to predict the price of a stock index (or stock price of a company) based on trends of the historical price of it \cite{picasso2019technical}. The Deep Learning methods play an important role in stock (index) prediction. They are used to train the information of historical patterns. This trained model can predict stock prices in the future. However, there are many different methods of how to train models, and how to select informative factors as inputs that are also stock index such as gold, crude oil, etc \cite{jimenez2020feature}.

To the best of our knowledge, no existing works have studied the cointegration-test and quantile loss simultaneously in stock index prediction using deep learning.  Firstly, A vast majority of works have not focused on how to select factors to train a model as input data in the stock prediction. Especially, no earlier works have studied that using cointegration test to select meaningful factors can provide better performance in terms of returns than in the case it is not used. In the TA, to predict stock index, it is important to select different factors well that are stock (index) prices related to factors we will predict such as s\&p500, etc. Next, as recent studies point out, the performance of deep learning models depends partly on which loss function you choose, such as quantile loss, rmse loss. In most regression tasks, mean absolute error and mean squared error are used popularly to train deep learning models. However, the quantile loss that we will reexamine in this paper gives better performance than the mse loss when evaluating it in terms of returns.

This paper aims to address that cointegration-test is excellent to select factors as inputs learning a model and also quantile loss is better than rmse loss that is commonly used to get higher returns.

 Our study’s key contributions are as follows:
\begin{itemize}
    \item We show that when building a deep learning-based prediction model to predict the stock index, factor selection method using cointegration test improves the performance of the model when evaluating the model in terms of returns
	\item We show that a regression model using Quantile Loss rather than the commonly used RMSE Loss gives better performance of the model in terms of returns
\end{itemize}

\section{Related Work}
\label{sec:headings}

\subsection{Cointegration Test}
Empirically, most financial time-series are non-stationary. Some researchers analyze whether there is a long-run equilibrium between this unstable financial time-series, and then perform stock prediction based on multiple regression using several informative factors as explanatory factors. In general, many of the researchers perform an Augmented Dickney-Fuller test\cite{dickey1979distribution}, a kind of unit root test, to determine whether the data we collect is stationary or not before performing a cointegration test. The ADF test is a kind of Dickney-Fuller test, which is the process of examining whether it is $\rho = 1 $ or $\rho < 1$ in expressions modified from the random probability model  $y_t = \rho y_{t-1}+\epsilon_t$ to $\Delta y_t = (\rho \text{-}1) y_{t-1} +\epsilon_t$.
The Cointegration test \cite{engle1987co} is performed after the Unit Root Test is done. We use the Johansen test \cite{dwyer2015johansen}, which is widely used among the variety of methods for Cointegration tests. Using this method, we can determine whether there is a long-term dependence relationship between time series without obtaining stationary via differentiation from non-stationary time series. Let's explore how to verify the existence of cointegration with the following VAR(2) model. Equation (1) follows the process below and results in (5).
\begin{equation}
    y_t=a_1 y_{t-1}+a_2y_{t-2}+\epsilon_t
\end{equation}
\begin{equation}
    y_t-y_{t-1} = (\alpha_1 - I)y_{t-1} + \alpha_2 y_{t-2} + \epsilon_t
\end{equation}
\begin{equation}
    \Delta y_t = (\alpha_1 - I)y_{t-1} + \alpha_2 y_{t-1} - \alpha_2 y_{t-1} + \alpha_2 y_{t-2} + \epsilon_t
\end{equation}
\begin{equation}
   \Delta y_t = (\alpha_1 + \alpha_2 - I) y_{t-1} - \alpha_2(y_{t-1}-y_{t-2})+\epsilon_t
\end{equation}
\begin{equation}
    \Delta y_t = \Phi y_{t-1} + \beta_1 \Delta y_{t-1} + \epsilon_t
\end{equation}
In Equation (5), it is interpreted that cointegration does not exist in time series if $\Phi=0$, and cointegration exists in time series if $\Phi \ne 0$. If the above test method determines whether the cointegration exists, it can be seen that a long-term dependence relationship between time series holds.

\subsection{Correlation}

As another method for factor selection, we adopt a method that utilizes Pearson Correlation \cite{jebli2021prediction} to determine the correlation between financial time series. The independent data $x_i$ for the dependent factor $y$ will have a correlation coefficient close to zero, and the positively correlated $x_j$ data will be close to one. In our work, we select five arbitrary $x$ with a high positive correlation for the dependent factor $y$ and use them as the learning data of the prediction model. The formula for obtaining Pearson correlation coefficients is shown below.

\begin{equation}
    \rho_{xy} = \frac{\sum_i (x_i - \bar{x})(y_i - \bar{y})}{\sqrt{\sum_i (x_i - \bar{x})^2}\sqrt{\sum_i (y_i - \bar{y})^2}}
\end{equation}

where $\rho_{xy}$ is the correlation coefficient, the closer to 1, the two factors $x,y$ will be positively correlated, and the closer to -1, the negatively correlated.

\subsection{Quantile Loss for Stock Index Prediction}
\label{sec:quantile}
The loss function is used as an indicator to evaluate how well the designed regression model fits the data. By finding the weights that minimize the value of the loss function, we can design a regression model suitable for the data. We try to adopt a model with higher performance by comparing RMSE which is commonly used in regression models and Quantile Loss.

The Quantile loss is a loss function for predicting intervals. Since linear regression cannot be used when the data do not follow a normal distribution, quantile loss is used to predict the Q-quantiles of the outcome factor. In particular, in the case of financial products, complete risk protection is essential to form a portfolio. For this purpose, it is effective to use a quantile loss that is robust to outliers and can more accurately predict the tail of the revenue distribution \cite{allen2009minimizing}. As \cite{allen2011quantile} showed that information can be effectively used in the overall distribution by using quantile loss, our study also uses it to prove that it is a better loss function in terms of return. 

\subsection{Backtesting}

Backtesting is a method of verifying the effectiveness of an investment strategy by testing it based on the predicted price information of the specific portfolio such as S\&P500. The position of the S\&P500 index in backtesting is based on the Buy-Sell-Hold trading strategy. This strategy is as follows: it buys stocks when the target price increases significantly, sells them when the target price decreases and holds them when the target price does not increase or decrease. The aim of this strategy maximizes the Cumulative return \cite{babiak2020deep} of the portfolio S\&P500 during the period of backtesting.

\begin{equation}
G^{QT} =\sum_{n=1}^{T} g_n
\end{equation}

\begin{itemize}
	\item $g_n:$ the Daily Return
	\item $1 \sim\ T:$ the Period of Backtesting
	\item $G^{QT}:$ the Cumulative Return
\end{itemize}

We need to study what is the Daily Return. With an S\&P500 index, we will analyze the $g_n$. Firstly, the function to set the position of the S\&P500 index in backtesting is called $F_{pos}$, and the log return of the S\&P500 index is called $r_n$. The $F_{pos}$ is defined as follows. The $V_D$ is the actual value of the S\&P500 divided by the predicted value.

\begin{equation}
F_{pos} =
\begin{cases}
    1 & \text{if $V_D$ \textgreater \: 0.03, then buy}\\
    -1 & \text{if $V_D$ \textless \: 0.03, then sell}\\
    0 & \text{otherwise, hold}
\end{cases}       
\end{equation}
~\\

Then, we can find the position value $S_{pos}$ for the S\&P500 using the position function $F_{pos}$ and $V_D$. In the formula below, the map applies $F_{pos}$ to each element of the vector in $V_D$.

\begin{equation}
S_{pos} = map(V_D, F_{pos})
\end{equation}
~\\
Next, the log return $r_n$ of the S\&P500 index is defined as follows,

\begin{equation}
r_n = ln(\frac{P_{last}}{P_{first}})
\end{equation}

\begin{itemize}
	\item $P_{first}$: the first value of S\&P500 index
	\item $P_{last}$: the last value of S\&P500 index
\end{itemize}
~\\
Using the position value $S_{pos}$ for the S\&P500 and the log return $r_n$, the daily return $g_n$ can be calculated as follows,
\begin{equation}
g_n = exp(S_{pos} \cdot r_n)
\end{equation}
~\\
Therefore, the Cumulative return \cite{babiak2020deep} of the portfolio S\&P500 is expressed as:

\begin{equation}
G^{QT} = \sum_{n=1}^{T} exp(S_{pos} \cdot r_n)
\end{equation}

\section{Methods}

We compare the traditional methods and suggested techniques in stock prediction to evaluate the benefits of factor selection and quantile loss introduced above. The general method takes all data as inputs for learning a model without filtering all the collected data and applies commonly used RMSE loss in a forecasting task using deep learning. Our proposed ideas are as follows. In figure 1, we present the illustration of this process. Firstly, Informative factors are selected to get the higher performance of the prediction model using two methods of factor selection that are the cointegration test and Pearson correlation ordering. Next, the model is trained. We use two architectures that are LSTM, GRU, and two-loss functions that are Quantile loss, RMSE loss. Finally, the trained model is evaluated using backtesting in terms of returns.

\begin{figure}
	\centering
	\includegraphics[width=16cm]{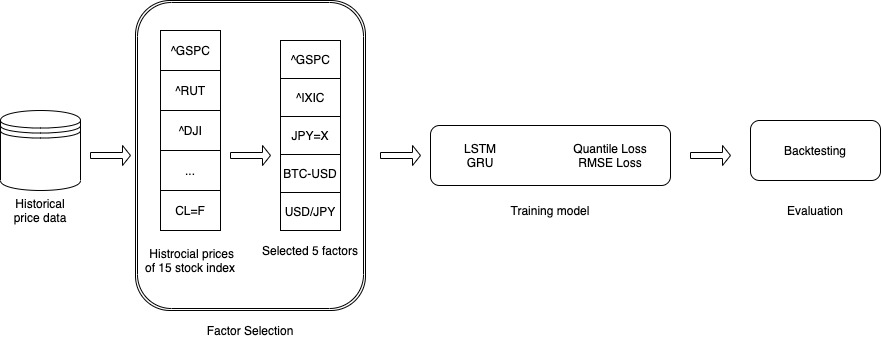}
	\caption{A Process of our proposed idea}
	\label{fig:process}
\end{figure}

\subsection{Model}
Time-series data has time-order information. In a methodology of the deep learning, to address this sequence data, it is general to use Recurrent Neural Network \cite{rumelhart1986learning}, \cite{werbos1990backpropagation}. Moreover, in the Stock Prediction task, it is popular to adopt the RNN architecture \cite{ozbayoglu2020deep}. The Recurrent model has the form $h_t=f(x_t,h_{t-1}; \theta)$, which means that it takes $x_t$ at a current timestep $t$ and $h_{t-1}$ hidden state at $t-1$ as input data and outputs $h_t$. It implies that the RNN memorizes the previous information to update current cell. However, this vanila RNN cannot capture long-term dependencies. To resolve this limitation, Long Short-Term Memory(LSTM) \cite{hochreiter1997long} and (GRU) Gated recurrent unit \cite{cho2014learning} are designed. Actually, both have been used frequently in the domain of stock prediction \cite{ozbayoglu2020deep},\cite{varaku2020stock}. In figure 2, we illustrate the mechanism of vanila RNN that is the foundation of LSTM, GRU in stock prediction.
The length of data is 601 ranged from 2018-03-05 to 2021-03-05. For the input $x_d$, d implies the day and $x_d = (s_{1,1},s_{1,2}, ... s_{5,p})$ where the $p$ is the number of selected stocks. The 5 implies the number of a weekday. At every timestep, the model predicts the S\&P500 stock index price of the next day. In the final layer, there is a Fully Connected layer $y_d = (k_1,... k_h)$ where the h is hidden size. Finally, the predicted stock index price $O_d$ in the future is output from the $y_d$.

\begin{figure}
	\centering
	\includegraphics[width=16cm]{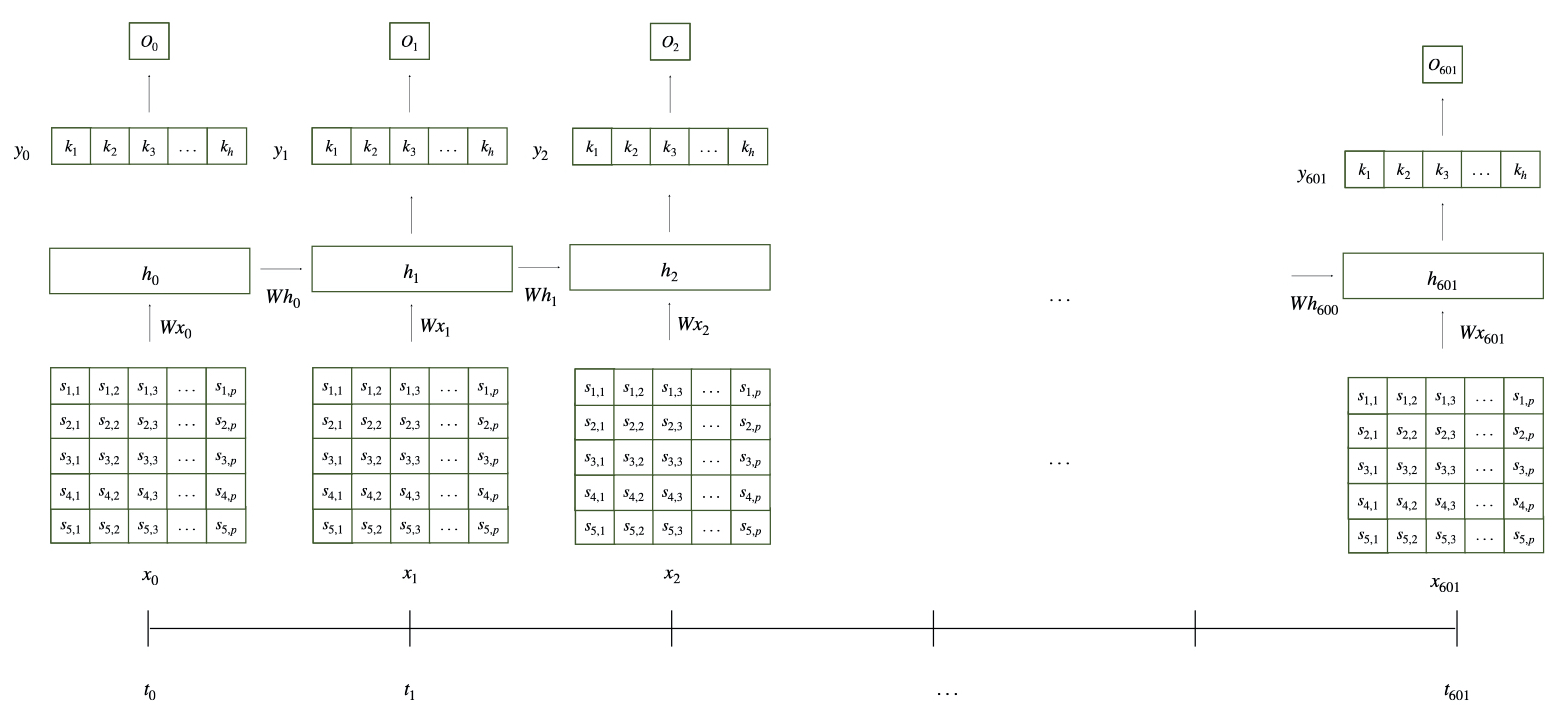}
	\caption{Architecture of the RNN model for stock index prediction }
	\label{fig:architecture}
\end{figure}

\subsection{Loss}
In this study, the quantile loss is calculated using a total of three quantiles: 0.1, 0.5, and 0.9. The average of the calculated loss of each quantile becomes the final quantile loss of the entire model. The definition of the calculation formula is as follows.

\begin{equation}
L_i^q = max((q-1)(y_i-\hat{y_i}),  q(y_i-\hat{y_i}))
\end{equation}
\begin{equation}
L = \frac{1}{n}\sum_{i=0}^n{L_i^q}
\end{equation}

\begin{itemize}
	\item $q$: a quantile
	\item $\hat{y_i}$: i-th predicted value
	\item ${y_i}$: i-th actual value
\end{itemize}
~\\

\subsection{Evaluation}
To evaluate the performance of our trading strategy for the portfolio S\&P500, we use two performance metrics that are Cumulative returns \cite{babiak2020deep} and Sharpe ratio \cite{jobson1981performance} \cite{memmel2003performance} \cite{leung2006testing}. Thus, we can evaluate the portfolio S\&P500 by comparing the two performance metrics of the portfolio S\&P500 index that we predict using the model and the SPY index that is the S\&P500 index in the actual world. In the Experiment, we compare both the Cumulative Return and the Sharpe Ratio between the portfolio S\&P500 index (12) and the SPY index. The Cumulative return of $G^{SPY}$ can be obtained as follows,
\begin{equation}
G^{SPY} = \sum_{n=1}^{T} exp(r^{SPY})
\end{equation}
\begin{itemize}
	\item $r^{SPY}$: the log return of SPY index
\end{itemize}
~\\
We can evaluate the portfolio by comparing the Cumulative 
Return, which is the value of $G^{QT}$ (12) and $G^{SPY}$ (15).

The second performance metric, the Sharpe ratio, can be expressed as follows,

\begin{equation}
Sharpe \: ratio = \frac{Expected\ Portfolio\ Return - Risk\ Free\ Rate}{Portfolio\ Standard\ Deviation}
\end{equation}
~\\
Let the Sharpe ratio of the portfolio S\&P500 be $S^{QT}$ and the Sharpe ratio of SPY be $S^{SPY}$. They can be expressed as

\begin{equation}
S^{QT} = \frac{r^{QT}}{\sigma^{QT}},\ \ 
S^{SPY} = \frac{r^{SPY}}{\sigma^{SPY}}
\end{equation}
As well as the Cumulative return, you can evaluate the portfolio with the value them.\\

\section{Experiments}
\subsection{Data}
To predict the value of the S\&P500 index in the future, we selected 15 stock index candidates for inputs and then trained the model using them or some of them picked by the method of factor selection. These time series were collected from March 5th, 2018 to March 5th, 2021. All of the collected data were used as inputs in the case that we perform a conventional method to predict stock index in the future, but when we do the suggested methods such as coordination test and Pearson-correlation to select important factors we use only 5 stock indexes of them. 

All data were collected from Yahoo Finance except for the U.S Federal Funds Rate. We collected Close Price data from March 15, 2018, to March 15, 2021. We processed missing values found in some indexes using interpolation. To perform a cointegration test, we do the ADF test on each index to identify whether all factors (stock indices) are stationary. The table below is the results.

\begin{center}
\begin{tabular}{ |p{3cm}||p{3cm}|p{3cm}|p{3cm}|  }
 \hline
 \multicolumn{4}{|c|}{Results of stock selection methods} \\
 \hline
 Index List & Ticker & ADF-test (p-value)& Cointegration test (p-value)\\
 \hline
 snp500   & \^\ GSPC  &0.742&0.000  \ \\
 Russell2000 &  \^\ RUT  & 0.776  & 0.888 \\
Dow Jones Industrial Average &\^\ DJI & 0.299 &  0.680\\
 NASDAQ    & \^\ IXIC & 0.946&  0.102\\
 USD/JPY &   JPY=X  & 0.093 & 0.160\\
 Bitcoin USD& BTC-USD  &  1.000  & 0.061\\
 FTSE 100 & \^\ FTSE  & 0.381 &0.508\\
 Nikkei 225 & \^\ N225  & 0.766 &0.740\\
 Treasury Yield 10 Years & \^\ TNX  & 0.911 &0.427\\
 EUR/USD & EURUSD=X  & 0.694 &0.647\\
 Gold Aug 21 & GC=F  & 0.798 &0.518\\
 Silver Jul 21 & SI=F  & 0.144 &0.700\\
 GPB/USD & GBPUSD=X  & 0.148 &0.491\\
 U.S Federal Funds Rate & \   & 0.909 &0.229\\
 Crude oil & CL=F  & 0.348 &0.773\\
\hline
\end{tabular}
\end{center}

\subsection{Method}
We compare all of the above factors as inputs to examine which model is best. However, in the case that uses Pearson Correlation, we select only five factors as inputs where we chose them in the highest order of correlation coefficient. On the other hand, in the case that uses the Cointegration test, we select five factors in the lowest order of p-value. Furthermore, we compare the utility of two using rmse loss and quantile loss respectively to see whether the model performs better in terms of returns in the case that we do not use rmse loss which is frequently used in regression tasks.
Finally, we do this experiment by adopting the most popular advanced RNN architecture designed to predict the price in the future of snp500. The table below shows the results of the experiment.

\begin{center}
\begin{tabular}{ |p{5cm}||p{3cm}|p{3cm}|p{3cm}| }
 \hline
 \multicolumn{3}{|c|}{Results of backtesting} \\
 \hline
   & Cumulative return & Sharpe ratio\\
 \hline
 All+Quantile+LSTM   & 4.304  &0.145 \\
 All+Quantile+GRU   &  4.397 &0.147 \\
 Correlation+Quantile+LSTM   & 26.781  &0.314 \\
 Correlation+Quantile+GRU   & 28.179  & 0.319 \\
 Cointegration+Quantile+LSTM   & 33.491  & 0.339 \\
 Cointegration+Quantile+GRU   & 36.779  & 0.351 \\
 All+RMSE+LSTM   & 2.382  & 0.089 \\
 All+RMSE+GRU   & 2.742  & 0.100 \\
 Correlation+RMSE+LSTM   & 11.171 & 0.230 \\
 Correlation+RMSE+GRU   & 12.963  & 0.245 \\
 Cointegration+RMSE+LSTM   & 3.838  & 0.128 \\
 Cointegration+RMSE+GRU   & 3.838  & 0.128 \\

\hline
\end{tabular}
\end{center}

In the above table, the `All' is the case that we use all given data. The `Quantile' refers to the case that we use quantile loss, and the `LSTM' also refers to the case that we use LSTM architecture. `All+Quantile+LSTM' is the combination of these cases. `Correlation' and `Cointegration' are also the cases we use Pearson Correlation ordering and Cointegration test as introduced above to select informative factors. 

\subsection{Discussion}
Analyzing based on the results of the above experiments, we can obtain the following insights: First, using factor selection methods gives us better performance in terms of returns than using all the collected data. Next, in our experiment, using the Cointegration test for factor selection gives us better performance than using correlation analysis. Third, using Quantile loss gives much better performance than using RMSE loss. Finally, in building a prediction model, LSTM and GRU do not make significant differences in performance.

In fact, in the case of the Cointegration test that tests the long-run equilibrium relationship between two factors, if the p-value is more than 0.05, some statisticians think that it can result in that the two factors do not have a long-term equilibrium relationship. However, in our experiment, we selected only five factors with the lowest p-value among them and used them as input values, even if the p-value of the cointegration test between the snp500 and all the remaining factors exceeds 0.05. Nevertheless, we suggest that the cointegration test can be a good choice for building a better prediction model for the stock index, given that the cointegration test used this test to input the selected factors, although the p-value of the test is higher than 0.05.

\subsection{Limitation}
Our study might have restraints. We need to point out the possible limitations of this study. Firstly, the 15 stock indexes used in this experiment can be a factor depending on the researcher's decision. The settings of the number and components of explanatory factors for training models can vary widely. For example, the number of inputs data used in \cite{hiransha2018nse} and \cite{shakhla2018stock} differs from that of the explanatory factors adopted in our study. Next, the experimental results may vary depending on the length of the data. The data used in this experiment are collected from March 5th, 2018 to March 5th, 2021. The results of the cointegration test with snp500 and the remaining explanatory factors collected during this period do not show that p-value < 0.05. On the other hand, the results of it with snp500 and remainings collected over a longer period show that the relationship between the Dow index and the snp500 is long-term equilibrium (p < 0.05).

\section{Conclusion}
The target of this paper was to show that a factor selection method and Quantile loss, which have not been highlighted in the stock index prediction task using deep learning, can be used to build the prediction model with better performance in terms of returns. We found that using factor selection methods such as the Cointegration test makes the model better performance in terms of returns than using all the stock index data collected in this work.
Furthermore, in this Stock Index Prediction, we found that using Quantile Loss would return better returns than using RMSE Loss, which is often used in regression or prediction tasks.
In the future, we will further explore whether other architectures such as Graph Neural Networks, Transformers, etc can perform better than the variant RNNs used in this work.

\end{document}